# 3D-mesostructures obtained by self-organization of metallic nanowires

**G.K. Strukova[1]\*, G.V. Strukov[1], S. V. Egorov[1], A.A.Mazilkin[1], I.I. Khodos[2], S.A. Vitkalov[3]**

*1 Institute of Solid State Physics, Russian Academy of Sciences, 142432, Chernogolovka, Russia*

*2 Institute of Microelectronics Technology and High Purity Materials, Russian Academy of Sciences, 142432, Chernogolovka, Russia*

*3 Physics Department, City College of City University of New York, NY 10031, USA*

\*E-mail: strukova@issp.ac.ru

## Abstract

The architecture of novel metallic mesostructures obtained via self-organization of growing nanowires has been investigated. Seashell-, fungus- and lotus leaf-shaped structures are reproducibly formed by programmable pulse current electrodeposition on porous membranes. The samples several millimeters in size are obtained. SEM investigation has revealed that the frame of the metallic "seashell" presents a hierarchical system with elements of fractal self-similarity at the nano- and micro-levels. The frame is a volumetric multilayer net with conical bundles of nanowires as building blocks. The Pd-Ni nanowires have V-like branches and periodic bulges ("beads"). TEM study showed that the nanowires consist of nanocrystallites dispersed in an amorphous matrix. Their sizes range from 4 to 15 nm. Local inhomogeneity of Pd-Ni solid solution was observed. In perspective, the proposed technique can be used as a 3D printer for the purposeful synthesis of novel materials with complex quantum nano-architecture.

## 1. Introduction

Mesoscopic hierarchical structures and methods of their fabrication are progressing rapidly due to the expected novel physical properties, effects and phenomena in materials based on them [1]. Recently we demonstrated an effective method for the fabrication of mesostructures by pulse current electrodeposition of metals from electrolytic solutions onto membranes with open pores. Such structures occur as a result of the self-organization of nanowires growing from membrane pores. One interesting property of the obtained mesostructures was their apparent similarity with natural objects: plants, shells, mushrooms (biomimetics) [2, 3]. Various micromodels of the plants, including hollow tubular containers with 10-20 nm thick walls, were grown by means of changing pulse current parameters, using different membranes and layer-by-layer electrodeposition from two different electrolytes [3]. Conditions for the reproducible growth of lotus leaf- and shell-like convex-concave microstructures were found [2-4]. The present paper describes the architecture and structure of the Pd-Ni alloy "seashells" grown by pulse current electrodeposition. The investigation may provide a better understanding of the leading mechanism of the growth of the potential quantum mesostructures.

## 2. Experimentation

Seashell-, fungi- or lotus leaf-like convex-concave structures were grown using a simple two-electrode circuit for the cathode electrodeposition of metals from an aqueous electrolyte. The electrolyte for the Pd-Ni electrodeposition contained (g/l): $PdCl_2$ - 6,0; $NiCl_2 \cdot 6H_2O$ -130,0; $NH_4Cl$ -75; ammonium sulfamate $NH_4SO_3NH_2$ -100,0; $NaNO_2$ - 30; aqueous solution $NH_3$ – up to pH=8. The Pd-Ni alloy electroplating was carried out in the temperature range 35-50 $^0$C. The cathode was a brass net covered with a polymer membrane 30 μm thick with an irregular pore diameter of 50-100 nm. The net size was $10 \times 30$ mm$^2$, the diameter of the wires was 100 μm, and the mesh size was $150 \times 150$ μm. A Pt plate served as anode. The cathode and anode were connected with a programmable current pulse generator. The structures shown in Figs. 1 and 2 were obtained at the following electrodeposition parameters: current amplitude in a rectangular current pulse was 100 mA; current pulse duty cycle, d.c. = 100 Ton/(Ton + Toff) % = 50 %, where Ton is the current pulse length and Toff is the pause between following pulses; the pulse frequency was 100 Hz; deposition time was 600-1000 s. The samples obtained were studied using SUPRA 50VP and JEOL scanning electron microscopes. The elemental composition of the samples was studied by X-ray spectral analysis using a scanning electron microscope SUPRA-50 VP. The "seashells" architecture was studied by comparing their images with those of fragments obtained by mechanical fragmentation or chemical dissolution (etching). The more detailed structure of the Pd-Ni samples was investigated using a JEM-2100 scanning electron microscope.

## 3.      Results

### 3.1.      General view

Electrodeposition of Pd-Ni alloy on membranes with disordered pore structure produced diverse "plant" structures (resembling "cauliflower", "broccoli", "squash", and "cactus") as well as convex-concave lotus leaf-, fungi- and shell-like structures [3]. Fig. 1 shows the "seashells" that are reproducibly obtained during the electrodeposition. The resultant convex-concave mesostructures can somewhat differ in shape owing to differences in the membrane geometry, pulse current mode or electrolyte composition. For instance, similar conditions enabled growth of lotus leaf-, fungi-, and shell-like microstructures from Pd, Ni, Pd-Ni, Pd-Co, FeNi$_3$, Cu and "cabbage-leaves" from Ag [2,3].

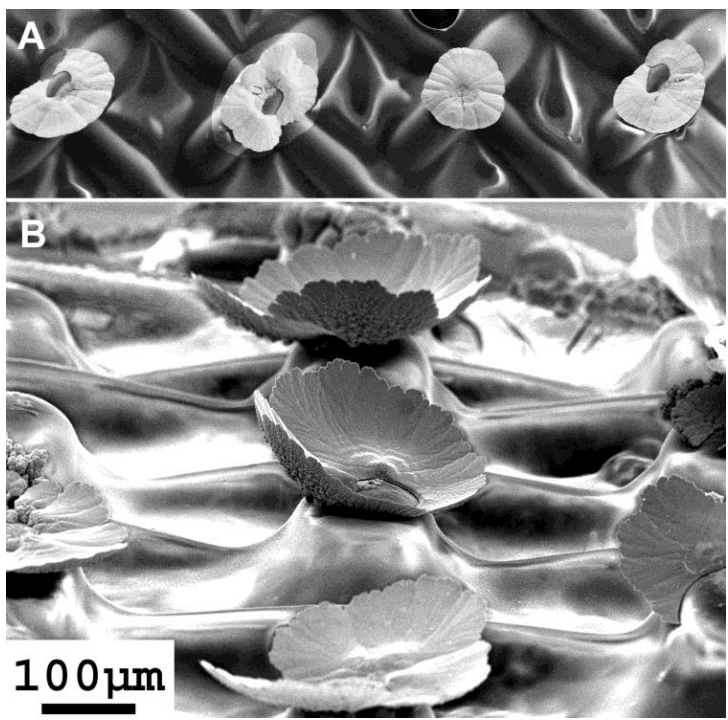

Fig.1. Convex-concave mesostructures (« seashells »): FeNi$_3$ (a); Pd-Ni alloy (b).

Pd-Ni alloy "seashells" with different Pd contents were obtained by electrodeposition from above-mentioned electrolyte using different current density, at that an increase in current density leads to an increase in the content of nickel in the Pd-Ni alloy. Fig.2 shows a Pd-Ni "seashell", alloy contains 25 at.% Pd. The mesostructure has the shape of a bell or cup.

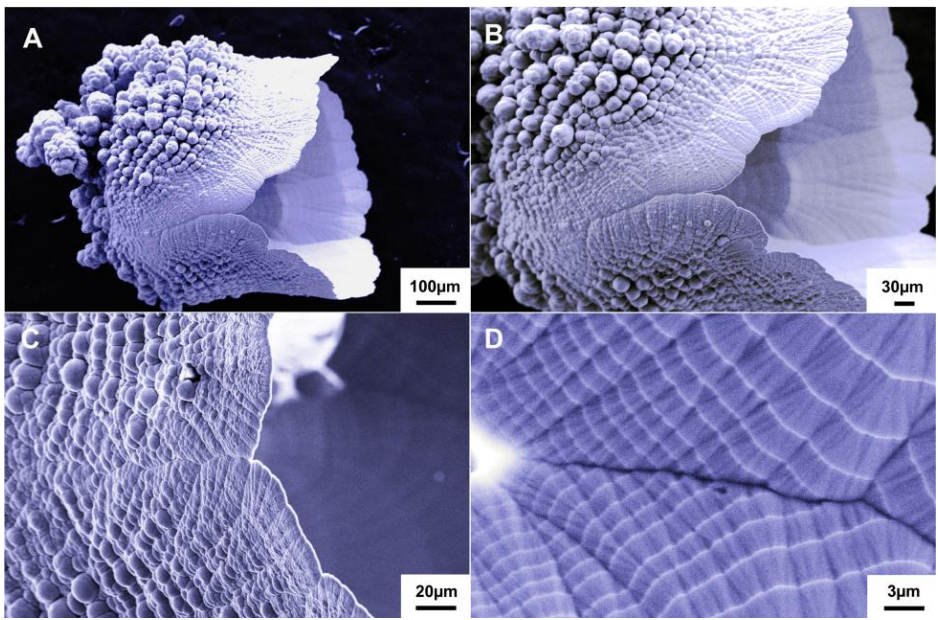

Fig. 2. Pd-Ni "seashell": general view (a,b), outer © and inner (d) surfaces.

These structures exhibit several characteristic features. The outer "shell" surface is often covered with "nanoflowers" that serve as templates for the nanowire growth (Figs. 2, 3).

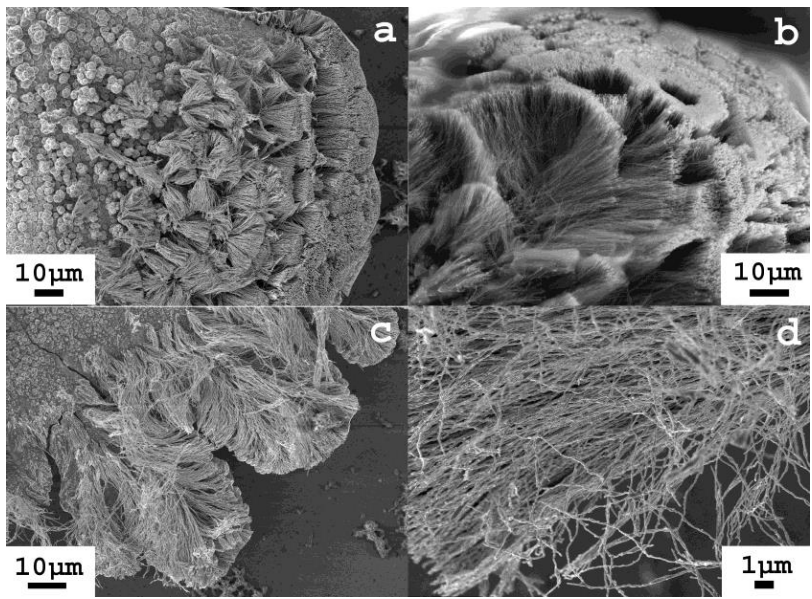

Fig. 3. Outer "shell" surface with "nanoflowers" and nanowires (a, c). View of nanowires (b, d).

The inner "shell" surface exhibits a characteristic pattern with lines going from the "root" to the periphery as well as with transversal circles (terraces) (Fig. 2).

## 3.2.   "Shell" microstructure

The grown mesostructures resembling "lotus leaves" and "shells" are examples of woven multilayer metallic materials. The surface pattern reveals the inner architecture of the "shell". The architecture was more exposed when the surface layer had been further removed by a mechanical treatment in an ultrasonic bath and/or by means of chemical etching. Fig.4a shows a "seashell" section with a partially removed inner surface layer.

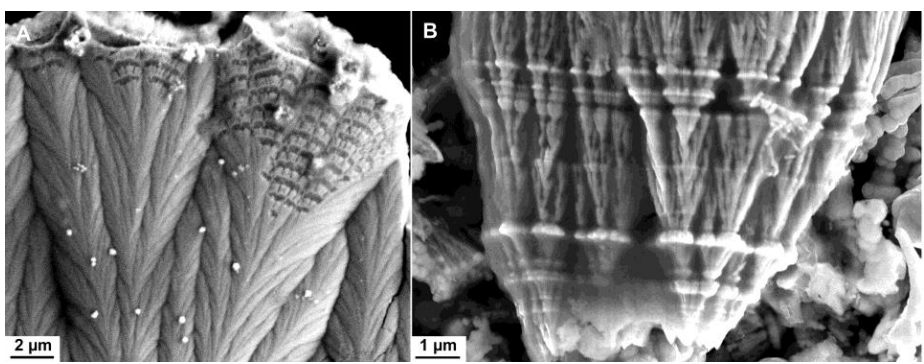

Fig.4. Pd-Ni "seashell" after removal of an inner surface layer (a) and after further treatment in ultrasonic bath (b).

The figure shows that the "shell" wall is composed of densely packed "bricks" that are conical bundles of nanowires. The bundle ends form "hoops" revealed on the inner "shell' surface as transversal circles or terraces (Figs. 2, 4). Fig. 4b shows another section obtained by a fragmentation of the "seashell" in an ultrasonic bath. A hierarchical system and a fractal self-similarity are visible here: the branching of the 100 nm conical bundles ("bricks") forms similar conical elements building the "seashell" wall but at a significantly larger (micrometers and tens of micrometers) scale. The images of the metallic seashell after fragmentation in the ultrasonic bath show that the tapered bundles of nanowires ("bricks") are packed as "seashell" wall layers (Fig. 5). Single nanowires separated from the bundles are also seen.

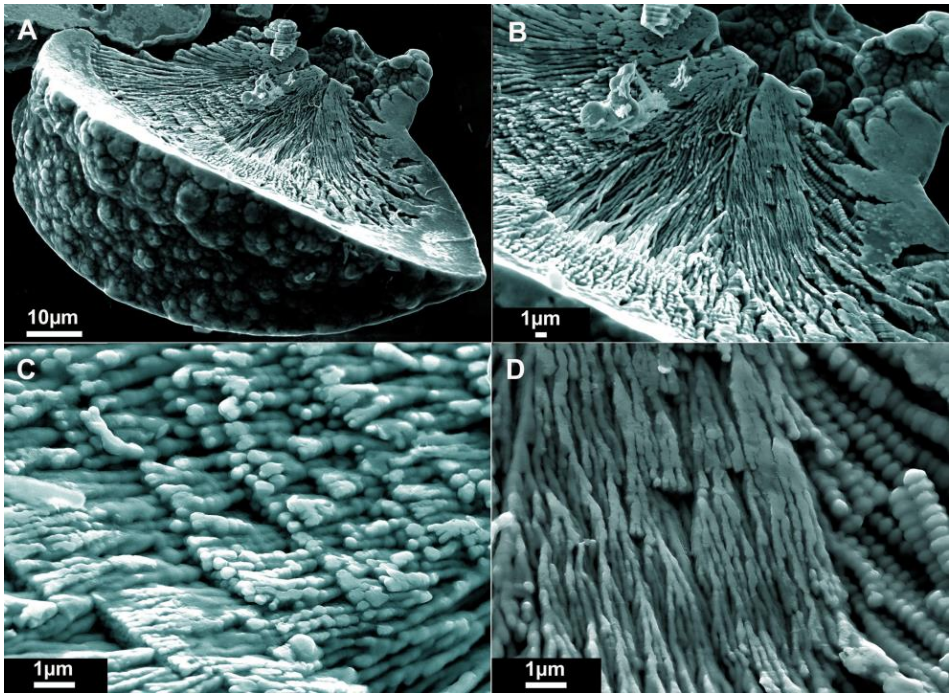

Fig. 5. Pd-Ni "seashell" and its parts after treatment in ultrasonic bath and chemical etching.

The hierarchical system, fractal branching and self-similarity are well seen in the images of the inner surface of the "seashell" after the chemical etching (Figs. 6, 7).

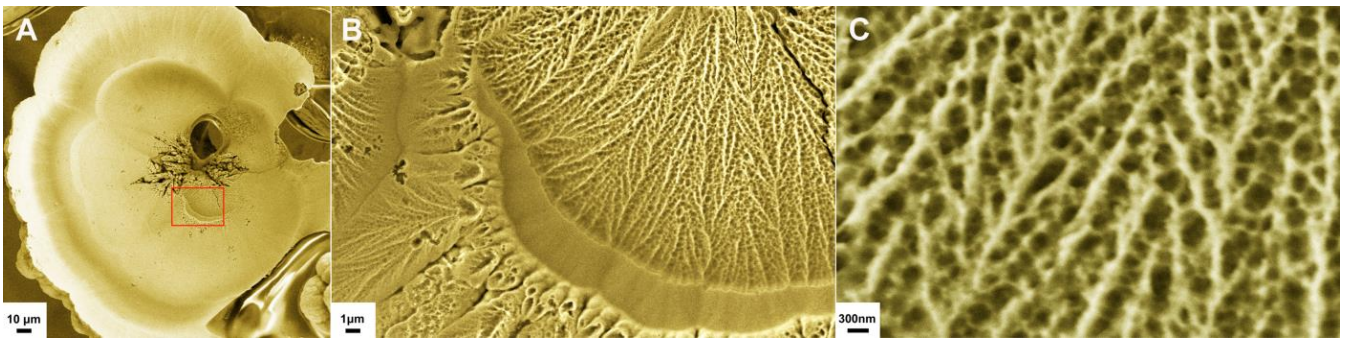

Fig. 6. Pd-Ni "seashell" before and after chemical etching.

These images suggest that the frame of the metallic seashell is a volumetrical multilayer net with conical bundles of nanowires as building blocks.

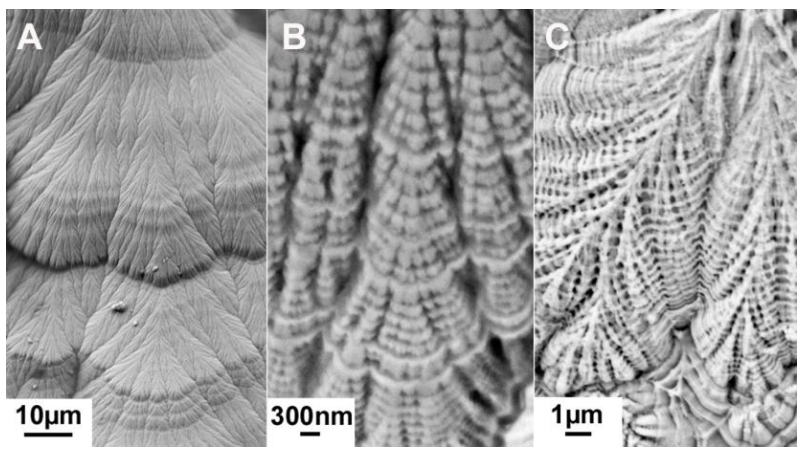

Fig. 7. Inner surface of Pd-Ni seashell: woven pattern on the surface (a); surface after 10 s etching (b); after 30 s etching (c).

Fig. 8 shows images of the metallic seashell after ultrasonic bath treatment, its fragments and single nanowires.

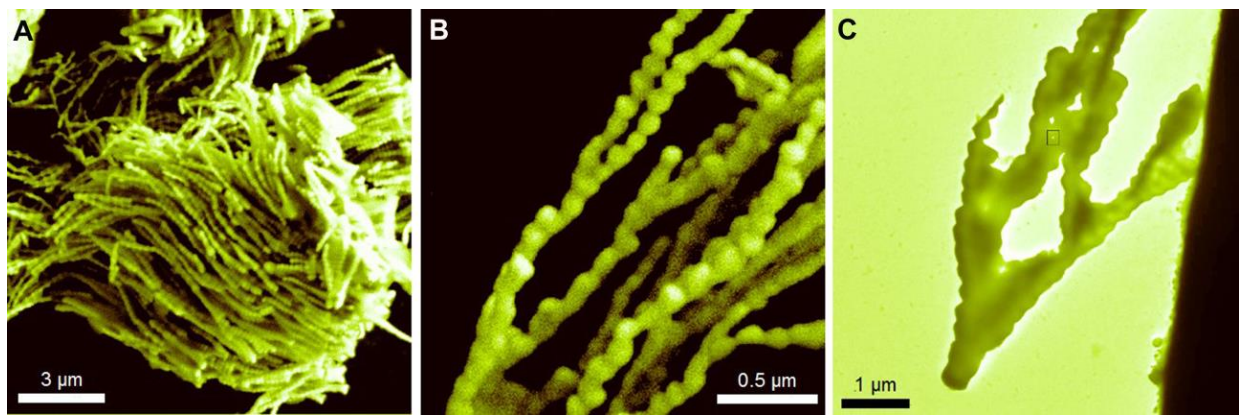

Fig. 8. Nanowires obtained after fragmentation of Pd-Ni "seashells" in ultrasonic bath.

It is seen that the "seashell" and all fragments consist of nanowire sections. The nanowires exhibit V-like branching and periodic bulges ("beads").

### 3.3. Elements of nanostructure

The structure of the metallic "seashell" and its fragments was studied in more detail using a JEM-2010 transmission electron microscope. The results of the TEM investigation reveal that the "seashell" nanowires consist of 4-15 nm nanocrystallites dispersed in an amorphous matrix. Fig. 9 shows images of a V-shaped nanowire with beads obtained after Pd-Ni shell fragmentation in an ultrasonic bath. Electronic diffraction pattern and high resolution image show the presence of the amorphous phase and nanocrystals. The radial distribution of the intensity of the diffraction pattern vs the scattering vector suggests the presence of two fcc crystal phases with cell parameters of 0.363 and 0.384 nm. (Fig. 9c).

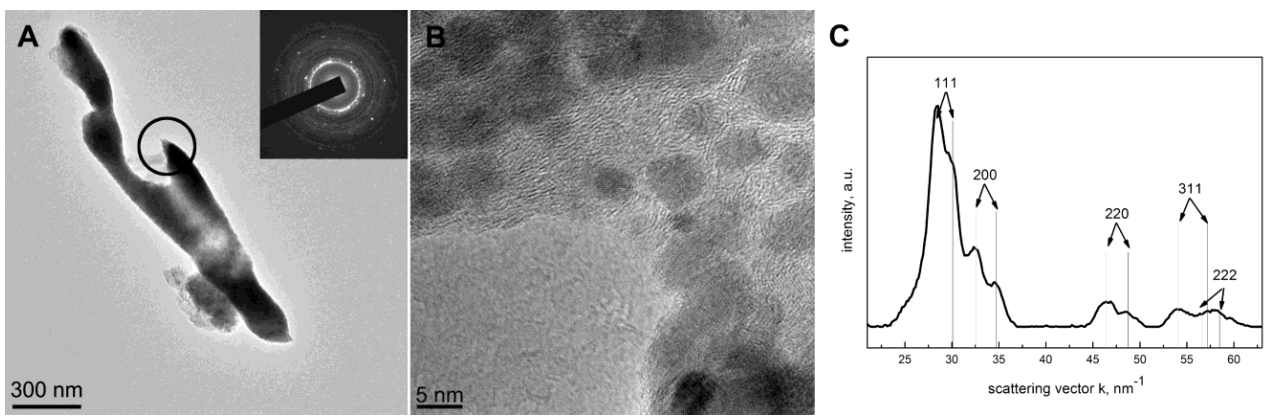

Fig. 9**.** Pd-Ni seashell nanowire: (a) general view; selected area of electron diffraction (SAED) is circled; electron diffraction of the selected area is shown in the insert to (a); a magnified high-resolution image of the selected area (b); plot of the radial distribution of intensity of the SAED pattern vs. scattering vector (c).

Also, thinned by chemical etching, the area near the root of the "seashell" revealed nanocrystalline inclusions embedded into a quasi-amorphous matrix.

## 4. Discussion

The reported results indicate that the programmable growth of convex-concave structures is facilitated by the combination of the shape of template and the pulse current mode. The template for these structures has pores which are shaped as a circle, oval or an open line (like horseshoe). Pores can be considered as "roots" of the growing structures. The pores are formed on portions of the membrane located above the crosshair of the net, where the upper wire forms a hill (Fig.1). Flat or tapered bundles of nanowires grow from growth points located on the edges of a "root". Such bundles growing from single points are frequently observed during the pulse current electrodeposition on the templates [3]. These structures do not occur during pulse current electrodeposition onto a smooth surface or during dc electro-deposition on the templates. These bundles (Figs.5,6) result from self-organization of branching bead-like nanowires (Fig. 8). The apparent advantage of pulse current electrodeposition over dc electrodeposition is the feasibility of fast growth at higher current density. This allows for deposition of metal both in the amorphous and in nanocrystalline phases, providing significant flexibility in the creation of the desired nanostructure [5]. In particular the nanowires, building the Pd-Ni seashell (see Fig.9), contain an amorphous phase with nanocrystalline inclusions of Pd-Ni solid solution. The nanocrystallites have 4-15 nm in size.

As shown in Fig. 2, the nanowire ends form rows and/or a network on the "seashell" surface. Such "nanoflowers" can serve as templates for new "nanoflowers" or nanowires (Fig. 3). The ends of nanowires and their surface roughness serve as templates for the growth of the frame of the "seashells". This is well shown in Figs. 4 and 5. Due to a singularity in the distribution of the electric field, the inner "seashell" surface is overgrown with a relatively smooth, reflecting film (Figs. 1, 2, 6, 7).

The proposed technique of template preparation can be significantly improved using modern methods based on photo and electron beam lithography. Designed arrays of structures with hierarchical nanoarchitecture with either identical or variable shape can be obtained. Additional investigations are required to understand the mechanism leading to the observed architecture of novel nanomaterials obtained in different pulse current modes. Furthermore, the proposed technique can be used as a 3D printer for the purposeful synthesis of mesostructures with complex quantum nanoarchitecture.

Described in this paper, the metallic mesostructures composed of nanowires are novel systems with complex architecture, including volumetric net with many angles and pores, composition and structural inhomogeneity. In accordance with contemporary understanding [1] such mesoscopic structures with hierarchical nanoarchitecture may exhibit novel macroscopic properties (transport, magnetic and optical) and effects providing an avenue for further development, applications and technologies based on quantum properties of matter.

Presented results suggest some promising directions for further investigations of both the physics of the mesostructure growth and the physical properties of the obtained metallic mesostructures. The "seashell" composed of ferromagnetic $FeNi_3$ nanowires (Fig.1) is an interesting object for investigating its interaction with a broadband microwave radiation since it is known that $FeNi_3$ nanoparticle conglomerates exhibit the ability to absorb microwave radiation over a wide frequency range [6]. Also, another interesting object for this direction is superconducting compound $Pb_7Bi_3$ that has been obtained in the nanocrystalline form by means of pulse current electrodeposition [7]. Very recently, different mesoscopic structures including mesoporous networks and nanowire mats attracted the attention of researchers as possible biosensors with high sensitivity [8, 9]. The convex-concave mesostructures composed of nanowires of precious metals (such as Au, Pd, Rh) and their corrosion-resistant and nontoxic alloys are good candidates for the using as biosensors. Considerable attention is being given to mesostructures with large surface area and to methods of the synthesis of these structures for solar energy applications [10, 11]. As presented in this paper, metallic mesostructures with complex architectural design could be valuable for applications in this field considering that the proposed method allows for the fabrication of metal- semiconductor composites using nanopowders such as CdS and $TiO_2$.

## 5. Conclusion

Novel types of nanostructured materials, nanowire-built metallic mesostructures, are obtained by the pulse current electrodeposition of metals from electrolyte solutions onto templates. Convex-concave structures are reproducibly fabricated via a self-organization of nanowires growth during the electrodeposition. The multilayer frame of these mesostructures presents a 3D network. The 3D network is a hierarchical system of conical nanowire bundles with elements of self-similarity and fractal branching. Nanowires of a bundle consist of nanocrystallites with sizes ranging from 4 to 15 nm dispersed in an amorphous matrix. The obtained metallic mesostructures with the hierarchical architecture, including volumetric net with many angles and pores, composition and structural inhomogeneity at the nano-scale are of significant interest for fundamental investigations of their physical properties. The projections based on "particle in a box" physical models indicate new physical properties of these hierarchical quantum systems, which are not attainable in accepted unquantized systems [1]. At the same time, the possibility of the direct manipulation of the architecture of mesostructures yields interesting prospects for creating novel materials with required physical properties. The mesostructures made of the normal, magnetic and superconducting metals and alloys could be valuable for applications based on the absorption of the electromagnetic radiation, in particular, for applications utilizing solar energy. Noble metal mesostructures of suitable size and shape can be grown and used as biosensors.

## 6. Acknowledgements

The authors thank colleagues from the Laboratory of Superconductivity—Anna Rossolenko, Ivan Veshchunov, and Valery Ryazanov—for their help in this work. We thank Roman Beletskiy for a useful discussion in the field of biosensors.